# BLINC: Lightweight Bimodal Learning for Low-Complexity VVC Intra Coding

Farhad Pakdaman[1], Mohammad Ali Adelimanesh[2], Mahmoud Reza Hashemi[2]

**Abstract** The latest video coding standard, Versatile Video Coding (VVC), achieves almost twice coding efficiency compared to its predecessor, the High Efficiency Video Coding (HEVC). However, achieving this efficiency (for intra coding) requires 31x computational complexity compared to HEVC, which makes it challenging for low power and real-time applications. This paper, proposes a novel machine learning approach that jointly and separately employs two modalities of features, to simplify the intra coding decision. To do so, first a set of features are extracted that use the existing DCT core of VVC, to assess the texture characteristics, and forms the first modality of data. This produces high quality features with almost no extra computational overhead. The distribution of intra modes at the neighboring blocks is also used to form the second modality of data, which provides statistical information about the frame, unlike the first modality. Second, a two-step feature reduction method is designed that reduces the size of feature set, such that a lightweight model with a limited number of parameters can be used to learn the intra mode decision task. Third, three separate training strategies are proposed (1) an offline training strategy using the first (single) modality of data, (2) an online training strategy that uses the second (single) modality, and (3) a mixed online-offline strategy that uses bimodal learning. Finally, a low-complexity encoding algorithms is proposed based on the proposed learning strategies. Extensive experimental results show that the proposed methods can reduce up to 24% of encoding time, with a negligible loss of coding efficiency. Moreover, it is demonstrated how a bimodal learning strategy can boost the performance of learning. Lastly, the proposed method has a very low computational overhead (0.2%), and uses existing components of a VVC encoder, which makes it much more practical compared to competing solutions.

## 1. Introduction

**VVC Standard:** The multimedia industry is always experiencing a dynamic change as new services and new technologies are introduced to customers constantly. Cloud gaming, virtual reality, ultra high-definition TVs, light field displays, and point cloud-based graphics are among these technologies. The Versatile Video Coding (VVC) standard [1] has been developed to facilitate such technologies and services, by improving the compression performance, which allows efficient communication and storage of huge video content. VVC has introduced several new or improved coding tools to achieve such coding efficiency. Some of these innovations are the large 128×128 pixels Coding Tree Unit (CTU), the Quad Tree with nested Multi-type Tree (QTMT) block partitioning, increased number of intra modes, new motion signaling modes, large transform cores of up to 64×64, and the Adaptive Loop Filter (ALF). VVC is reported to achieve a coding efficiency twice its predecessor, the High Efficiency Video Coding (HEVC) [2]. However, this higher efficiency comes at the price of several times higher computational complexity. For Low-Delay (LD) and All Intra (AI) configurations [3], VVC takes 5x and 31x the complexity of HEVC encoding and 1.5x and 1.8x HEVC decoding, respectively [4]. Such complexity, not only makes it difficult to deploy VVC for real-time encoding, even for video streaming, such codecs are cost-efficient only for videos with millions of views [5].

Although inter-frame coding is dominant in most video applications, intra coding is also very important on its own as (1) it is often used for image compression [6], it is used periodically to prevent error propagation and random access to video, and it has applications in complex networks [7], video surveillance [8], and video archiving [9]. VVC has invested specifically in improving the performance of intra coding, through providing more CTU partitioning modes for intra-coded CTUs, and extending the number of intra modes from 35 modes in HEV to 67 modes. As a consequence, all-intra VVC coding has become more complex compared to LD VVC coding (by 1.3x [4]), while it used to be much simpler than LD encoding in HEVC and previous standards. Therefore, it is essential to investigate complexity reduction techniques for VVC intra coding, in order to balance the complexity in VVC encoders, and also enable VVC encoding for all-intra applications such as image compression, archiving, and video surveillance.

**Baseline VVC intra coding process:** The baseline VVC intra prediction process, which is used in the VVC Test Model (VTM) [10], consists of exhaustively searching among the 67 intra modes, to find the best one. To remedy the complexity associated with the exhaustive search, VTM introduces a three-

Corresponding Author: Farhad Pakdaman (Email: farhad.pakdaman@umz.ac.ir)
[1] Faculty of Engineering and Technology, University of Mazandaran, Babolsar 4741613534, Iran.
[2] School of Electrical and Computer Engineering, College of Engineering, University of Tehran, Tehran 19466-14886, Iran.

step intra search. At step one, the intra search is performed using a Rough Mode Decision (RMD), which computes the coding cost based on the Hadamard cost. The angular modes with an even index are searched with RDM, to find the *R* best modes. Then, the two modes at either side of the best modes are checked. Second, the Most Probable Modes (MPM) are added to the list. Finally, the more complex Rate-Distortion Optimization (RDO) is performed on the list of best modes, to find the final best mode. Although this process reduces the complexity compared to the exhaustive search, it is still complex, and the search includes an unnecessarily large number of modes that could be avoided using a smarter algorithm.

**Machine Learning based complexity reduction:** Several researches have been dedicated to reducing the complexity of HEVC or VVC intra mode decision [11][12][13]. Among these works, Machine Learning (ML)-based methods achieve the best results, as they can learn through examples, considering several video properties, such that the learned model can generalize to new observed data as well. Support Vector Machines (SVM) [11][14], Convolutional Neural Networks (CNN) [12][15], and online training strategies [16] are some of these successful solutions. However, these techniques have some important drawbacks that need to be addressed: (1) most of the existing solutions are tailored for HEVC and are oblivion to the more detailed intra angular modes, and larger CTU size of VVC, (2) the ML-based solutions are often preceded by a complex feature extraction step, or use a complex inference structure such as CNN, which adds a considerable complexity overhead, making these solutions nonpractical, (3) each method uses a single source of data, i.e. data modality, for model training, such as texture characteristics, bitstream level symbols, or frame statistics, without benefiting from other modalities, and (4) either an offline training is used that cannot be updated according to the content, or an online training is used which adds a considerable overhead.

**The proposed method:** To address the above-mentioned challenges, this paper uses the following intuitions to designs an efficient and low overhead ML-based approach:

(1) The existing internal tools of VVC encoder are used for feature extraction to minimize the overhead. It is demonstrated in [17] how the residual of the HEVC Planar intra mode can efficiently be used to model the texture direction. With this intuition, the residual of three common modes, Planar, Vertical and Horizontal modes are used for feature extraction.
(2) Discrete Cosine Transform (DCT) is a strong transformation core which is a main part of all VVC encoders. As this core is used for the final encoding decisions and bitstream generation, it can also be used to analyze the texture characteristics of a block, using the intra modes mentioned above.
(3) Video encoding is expected to be carried out in real-time in many power-constrained devices. Hence, the key to a practical fast encoding is to learn the task with a lightweight and efficient model. Using Multilayer Perceptron (MLP) along with a reduced and compact feature space, can guarantee a lightweight and yet powerful learning algorithm.
(4) Several different features can represent the encoding decision, such as texture characteristics, distribution of modes in neighboring blocks, and bitstream level symbols such as coding flags. Multimodal learning enables fusion of information from more than one modality, to achieve a more robust algorithm.

Following these intuitions, this paper proposes a Bimodal Learning approach for VVC Intra Coding (BLINC), that effectively reduces the computational complexity. First, two sets of features are extracted that correspond to two modalities features, for the intra coding decision task. The texture modality is extracted using the DCT coefficients of specific intra modes. As this process is already part of any VVC encoder, this adds no overhead. The second modality is the distribution of intra modes in the neighboring area, which is obtained by simply recording the best modes of neighboring blocks. Second, a two-step feature reduction is proposed that transforms the features into a much more compact space, such that they still represent the same data. This allows the task to be learned using a simple and lightweight model, avoiding high overheads. At the third step, three different training strategies are proposed to learn the intra mode decision task. Offline and online training strategies are proposed that use a single data modality, i.e., the texture and the distribution of modes, respectively. A mixed online and offline training strategy is also proposed that uses a late fusion for a bimodal learning approach [18], to benefit from both data modalities and improve the training performance. Finally, a fast intra coding algorithm for VVC is proposed, based on the proposed training strategies, that achieves a promising complexity reduction, with a small loss of efficiency.

**Major contributions:** The major contributions of this paper can be summarized as follows:

(1) A feature extraction method with near-zero overhead is proposed, which is based on DCT coefficients and neighboring blocks. The features include two data modalities that correspond to the same event (same label).
(2) An offline ML strategy is introduced that learns the intra mode decision, with a lightweight and low overhead model.
(3) An online training process for fast intra mode decision is proposed, that efficiently trains on few samples of data, including the best modes of the neighboring blocks.
(4) A Bimodal ML strategy for complexity reduction of video coding is introduced for the first time, which is demonstrated to improve the learning performance above each unimodal model.

The rest of the paper is organized as follows. Section 2 discusses the related works. Section 3 details the proposed method. The experimental results are presented in section 4, and finally, section 5 concludes the paper.

## 2. Related Works

As the VVC standard has been recently finalized in July 2020, some papers have addressed its complexity, while many aimed to accelerate HEVC intra coding [19]. This section summarizes some of the successful solutions for complexity reduction of intra coding.

## 2.1 Fast CU partitioning

As the introduction of the new CTU structures, such as the Quad-Tree in HEVC and Quad Tree with the nested Multi-Type Tree (QTMT) in VVC has added significantly to the complexity of CU partitioning, many works have addressed this issue. Texture complexity has been known to have a strong correlation with CU partitioning decision. To measure this complexity, the pixel variance [20], the edge energy [21], and global and local texture energy [22] have been used for HEVC. Liu et al. [21] used the directional Sobel filters to measure the directional and sub-block complexities. Then, SVMs are trained to classify each block into a split or non-split block. Grellert et al. [23] use a similar technique for HEVC, using bitstream level features, such as the prediction modes, coding flags, average depths of neighboring CUs, and coding modes. A similar extended feature set is used in [24], to predict the CU partitioning for VVC, using the Light Gradient Boosting Machine (LGBM). Pakdaman et al. [11] proposed extracting a feature set using the dual-tree complex wavelet transform (DT-CWT), which outperform the pixel-level features. Then, a decision-making framework is presented using SVM and constrained optimization, that adaptively decides the CU partitioning according to the available processing power. Fast CU partitioning based on Just Noticeable Distortion (JND) [25] is another notable work [26].

Deploying CNNs and deep learning models for fast CU partitioning have recently been a hot topic [24][27][28]. Xu et al. [27] proposed a deep learning based fast CU partitioning for HEVC. A CNN based early termination decision model is presented that learns the spatial features for fast intra mode decision. Moreover, the correlation between CUs is learned using a Long Short-Term Memory. As the number of CNN layers for fast encoding increases, the propagation of features and gradients through the network becomes challenging. Zhang et al. [24] proposed an architecture based on DenseNet with skip connections, and attention mechanism, to remedy this issue. Using CNN to estimate the optimal multi-type tree partitioning is among the other notable efforts [26].

## 2.2 Fast intra mode decision

As HEVC and especially VVC support a large number of intra modes, researches in this category aim to decide the best intra mode, avoiding an exhaustive search. Gradient of luma samples has been used in several researches to estimate the best angular mode [13][14][29]. Zhang et al. [14] classify intra modes of HEVC into nine classes, and decide the best class of modes by comparing each block's gradient in horizontal and vertical directions, to the average gradient. Zhang et al. [30] propose a random forest classification for VVC intra modes, that also classifies all intra modes into four groups, using pixel-level texture complexity. Dong et al. [13] proposed a two-part mode decision and fast CTU partitioning method for VVC. First, the feasibility of the newly introduced Intra Block Copy (IBC) and the Intra Sub-Partitions (ISP) are analyzed for each block, as these tools are not suitable for all blocks. Features such as the number of pixels with large gradient, the maximum gradient magnitude, number of zero gradient pixels, and context information are used to decide the early termination of these modes. Second, an ensemble decision strategy is used to sort the candidate modes based on the Hadamard cost and the mode distribution. To achieve more representative features compared to pixel domain features, the use of DT-CWT and Prewitt operators have also been suggested in [31] and [32], respectively.

While the above-mentioned techniques can effectively model the intra mode decision, they add extra computation for feature extraction. To remedy this, Hosseini et al. [17] showed that the residual of HEVC planar intra prediction can effectively be used to estimate the edge direction. Combining this technique with a pareto frontiers optimization, authors propose a fast HEVC intra coding that sets the best number of modes to check, according to the available processing power [33]. Efficient hardware approximation of complex codec tools is among the other low-overhead approaches [34].

CNN-based techniques have also been proposed for fast intra mode decision. Laude et al. [12] proposed a CNN architecture for classification, consisting of two convolutional layers, a fully connected layer, and a SoftMax layer, that predicts the best intra mode. Chen et al. [15] also used CNN, but modeled the problem into a regression task, where the minimum number of intra mode candidates is determined according to the block's texture. Moreover, Tsang et al. [29] used a U-Net based solution to decide between natural intra content and screen content, in VVC screen content coding.

While the previous works discussed above can reduce the computational complexity of intra coding, all of them suffer from one or more of the following shortcomings: (1) most existing works target HEVC and much more efforts are yet required to achieve a solid solution for VVC, (2) the ML-based methods either require a heavy feature extraction step, or use CNNs that add a considerable overhead and cannot perform in real-time without a Graphics Processing Unit (GPU), (3) existing methods often use a single modality of feature, such as texture information or the distribution of intra modes in the frame, and do not benefit from both feature modalities, (4) existing works either use a rigid offline trained model, or only use an online model that adds considerable overhead. In the next section, the proposed ML-based solution is detailed, which alleviate the existing shortcomings by introducing a novel lightweight bimodal training strategy.

## 3. The Proposed Method (BLINC)

This section details the proposed low complexity VVC intra encoder, named BLINC. Fig. 1 depicts the overall framework of BLINC. As this work relies on ML, at the first stage (discussed in 3.1) a feature extraction method is presented that captures two modalities of data for the intra mode decision. The DCT coefficients of three intra modes, that represent the texture characteristics, are extracted as the first data modality. The best modes of neighboring blocks, which represents the distribution of intra modes in the frame, are extracted as the second modality. To be able to learn the intra mode decision task efficiently and with low complexity overhead, a high-quality feature space with limited dimensions is required. Hence, at the second stage (discussed in 3.2), a two-step feature reduction method is proposed that uses a selective max pooling followed by Principal Component Analysis (PCA). This effectively reduces the size of feature space from up to 12288 features (for a 64×64 block) to only 15 features for all block sizes. As it can be observed in Fig. 1, the third stage (discussed in 3.3 and 3.4) discusses training strategies to learn the intra mode decision task. Depending on the data modality, online and offline training strategies are proposed that use a lightweight shallow MLP to learn a unimodal feature space. Moreover, a mixed online/offline strategy is proposed that uses a late fusion to provide a bimodal learning. At stage four (discussed in 3.5), the encoding algorithm of BLINC is presented that uses the proposed learning strategies for fast and low-complexity VVC intra encoding. Table 1 summarizes the important notations used throughout the next subsections.

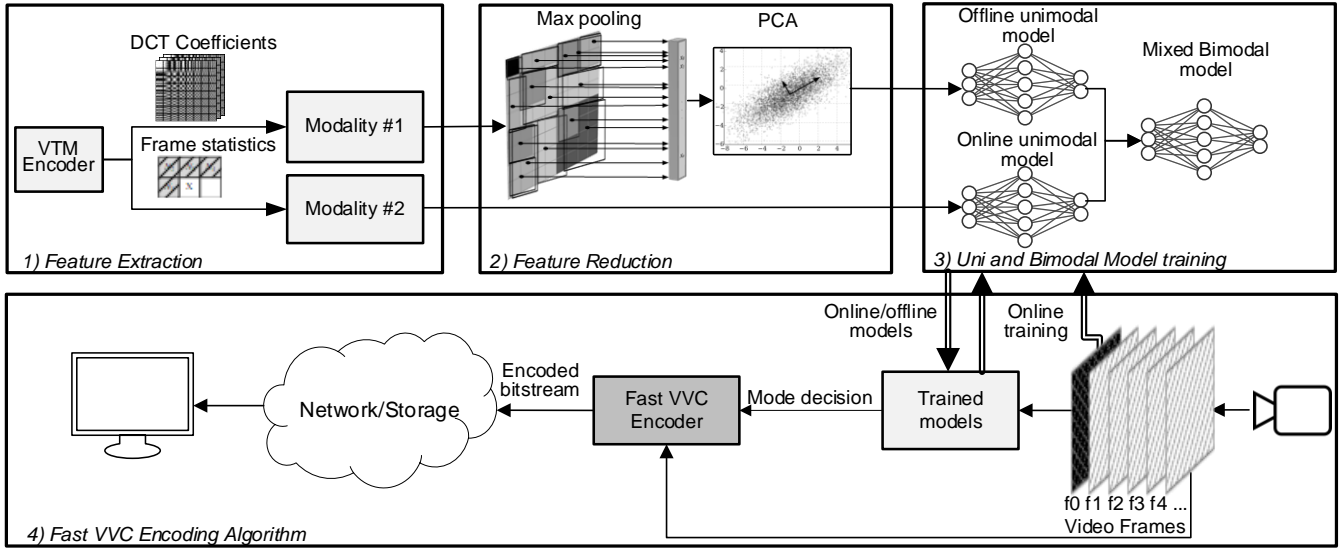

**Fig. 1.** Overall framework of BLINC

**Table 1.** Table of important notations used throughout the paper

| Notation | Definition |
|---|---|
| $B$ | Block of luma samples |
| $Pred_i$ | Prediction signal for intra mode $i$ |
| $Res_i$ | Residual signal for intra mode $i$ |
| $C_i$ | DCT coefficients of $Res_i$ |
| $m_i$ | Vector of max DCT values obtained from DCT max pooling |
| $x_1, x_2$ | Feature vectors of the first and second data modality |
| $W$ | PCA transformation matrix |
| $y \in \mathbb{N}^9$ | One-hot ground truth (label) for observed data $x$ |
| $\hat{y} \in \mathbb{R}^9$ | Predicted output |
| $\hat{y}_1, \hat{y}_2$ | Predicted outputs using the first and second modalities, respectively |
| $\hat{y}^{(i)}$ | Predicted output for class $i$ |
| $\theta$ | Parameters of the hidden layer, including weights and biases |
| $x_3$ | concatenated predictions of online and offline model |
| $K$ | Number of intra mode classes to check at the encoder |
| $T$ | Threshold for considering the intra mode score high |
| $R$ | Number of RDO mods to check |

### 3.1 Feature extraction for two modalities

The proposed method uses two sets of features, as two data modalities (two viewpoints or measures of the same event) to train ML models. This subsection introduces the feature extraction for these two modalities.

The first feature modality, $x_1$, represents the texture characteristics and is extracted from the DCT coefficients of intra prediction residuals. To do so, first the intra prediction is performed for intra modes 0, 18, and 50, which are the planar, horizontal, and vertical modes, respectively. As already discussed in section 1, these modes are the three most frequent modes, and their residual carries important information on blocks texture. Considering $Pred_i$ as the prediction of block $B$ using mode $i$, its associated residual signal is calculated next, via pixel-wise subtraction, as in (1).

$$Res_i = Pred_i - B \quad (1)$$

Next, the residual signal is decomposed using a 2-d DCT transform, to obtain $C_i$, the DCT coefficients as a matrix of the same size as $B$, as in (2). Subsection 3.2 explains how $C_i$ is reduced in dimensions to obtain $x_1$.

$$C_i = DCT(Res_i) \quad (2)$$

The second feature modality represents the distribution of intra modes in the vicinity of the current block, and consists of the Best Intra Mode (BIM) of four previously coded blocks. These blocks as suggested by (3), are the left, upper-left, top, and upper-right blocks. Fig. 2 shows these modes.

$$x_2 = \langle BIM_L, BIM_{UL}, BIM_U, BIM_{UR} \rangle \quad (3)$$

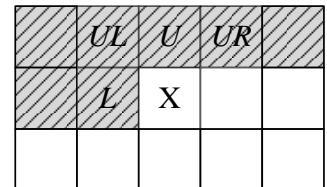

**Fig. 2.** Neighboring blocks selected for the second feature modality

Along with the two feature modalities, the best intra mode, $y$, is also extracted as the ground truth for each block. Fig. 3 shows the intra modes of VVC. As VVC supports 65 angular intra modes, learning a model to predict the exact intra mode would not be practical as (1) learning a model to predict 65 distributions requires a complex model and many training samples, and (2) as adjacent modes are very similar, learning to predict them with such a precision is often not possible and the trained model would not perform well. In fact, discriminating between similar modes is only possible by comparing their rates and distortions while encoding [31]. Hence, grouping adjacent intra modes is a popular trend in ML-based fast intra prediction [30][35][36]. Here, the adjacent modes are grouped into 9 classes and $y \epsilon \mathbb{N}^9$ is a one-hot code indicating one of the 9 classes of intra angular modes. These classes can be observed in Fig. 3. Subsections 3.2 and 3.3 discuss how the probability of each class is learned, and subsection 3.4 details how a list of candidate modes for the encoding process is derived, having the probability of each class.

calculated as (4) and (5), and visualized in Fig. 4. These coordinates are selected empirically to cover various frequency bands. Here, $C_i(x1, y1 - x2, y2)$ selects the coefficients between the two coordinates of (x1,y1) and (x2,y2) in matrix $C_i$. For other block sizes, the same selections expand or shrink, such that always 15 features are selected. As these neighboring coefficients correspond roughly to similar horizontal or vertical frequencies, this operation merges every few similar values into one representative feature.

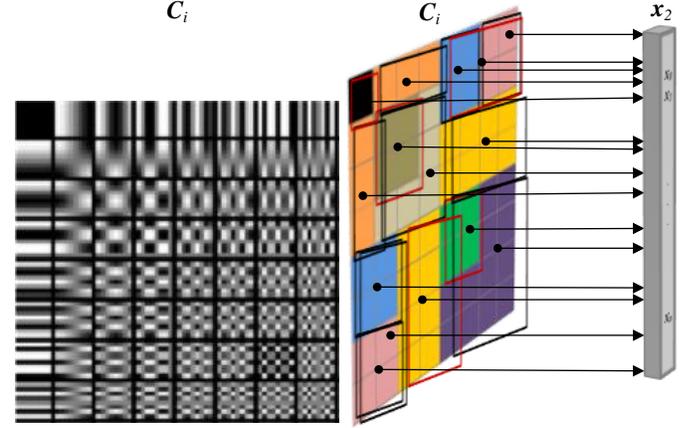

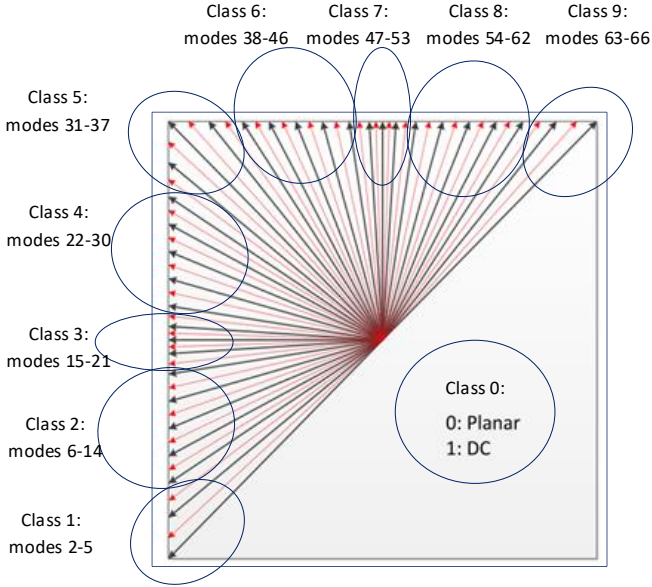

**Fig. 3.** VVC intra modes, grouped into 9 intra angular mode classes

**Fig. 4.** Left: representation of DCT coefficient as their frequency characteristics, right: representation of the selective max pooling to achieve 15 features, both for a block of 8×8

$$\boldsymbol{m}_i = \langle m0, m1, \dots, m14 \rangle \qquad (4)$$

$$\begin{aligned}
m0 &= |C_i(0,0)| & m1 &= \max |C_i(0,1-0,3)| \\
m2 &= \max |C_i(0,4-1,5)| & m3 &= \max |C_i(0,6-1,7)| \\
m4 &= \max |C_i(0,4-1,7)| & m5 &= \max |C_i(1,1-2,2)| \\
m6 &= \max |C_i(1,1-3,3)| & m7 &= \max |C_i(1,0-3,0)| \\
m8 &= \max |C_i(4,0-5,1)| & m9 &= \max |C_i(6,0-7,1)| \\
m10 &= \max |C_i(4,0-7,1)| & m11 &= \max |C_i(2,4-3,7)| \\
m12 &= \max |C_i(4,2-7,3)| & m13 &= \max |C_i(4,4-5,5)| \\
m14 &= \max |C_i(4,4-7,7)| & & \qquad (5)
\end{aligned}$$

### 3.2 Feature reduction

While the second feature modality, $x_2$, only consists of 4 elements, the first modality, $C_i$, contains the same number of elements as the original block. Considering three intra modes ($i$=0, 18, 50), for blocks of 8×8 or 64×64 pixels, this means 3×8×8=192 or 3×64×64=12288 features. While these features can effectively describe the texture characteristics, learning such a large feature space with many corelated features requires a model with a high capacity, and hence, a high computational overhead [37][38]. To be able to use a lightweight model, a two-step feature reduction scheme is presented here, that reduces the feature size into only 15 for all block sizes.

Firstly, to achieve the same number of features for all block sizes, and also to gain a coarser granularity of frequency coefficients, a selective max pooling is proposed, that finds the maximum value among the selected neighboring coefficients. For coefficients of mode $i$ corresponding to a block of 8×8 pixels, $C_i$, the vector $m_i$ consists of 15 maximum values,

Next, the three vectors $\boldsymbol{m}_0$, $\boldsymbol{m}_{18}$, and $\boldsymbol{m}_{50}$ that correspond to the three intra modes, are concatenated as in (6), to form a feature vector with 45 features. While this vector is much smaller than the initial $C_i$, it consists of components from three residual signals, which are correlated and hence, carry redundant information. Therefore, PCA is used to analyze these features and decorrelate them. PCA learns an orthogonal transformation matrix, $\boldsymbol{W}$, that transforms the data to a lower-dimensional representation, such that the data has the maximum variance along the new dimensions. After learning the matrix $\boldsymbol{W}$, the reduced form of the first feature modality can be calculated via PCA transform, as in (7).

$$\boldsymbol{m} = \boldsymbol{m}_0 || \boldsymbol{m}_{18} || \boldsymbol{m}_{50} \qquad (6)$$

$$\boldsymbol{x}_1 = \boldsymbol{m}^T \boldsymbol{W} \qquad (7)$$

Fig. 5 reports the variance of the training data over each new dimension (principal component), for the first 20 components out of 45 components. It is clearly observed that the variance for the new dimensions falls rapidly, suggesting that the first few components can effectively represent the data. To be consistent among all CU sizes, the first 15/45 features are selected for $x_1$.

At the end of this stage, two sets of features, $x_1$ for the first modality and $x_2$ for the second modality, as well as the class labels, $y$, are available that will be used for model training, in next subsection.

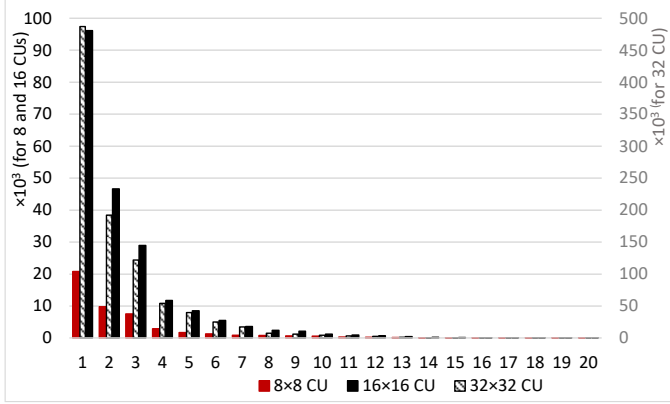

**Fig. 5.** Variance of the first feature modality, along the first 20 principal components. Components 21-45 have been removed as they are very small

### 3.3 Uni and Bimodal model training

Using the extracted feature sets in previous subsections, this subsection proposes three training strategies for learning the intra mode decision task. First an offline training strategy is introduced using the first modality of features. Second, online training is selected to learn the second modality for two reasons (1) as the distribution of intra modes (i.e., the second modality of data) varies for each video scene, offline training would not account for different video scenes, and (2) as frame statistics is the only information in this modality, using an offline trained model can make the encoder drift towards the dominant modes in the scenes over the time. Third, a mixed model is introduced that mixes the predicted scores of the two online and offline models for a bimodal prediction.

For all three training strategies, a lightweight MLP, with only one hidden layer is used. MLPs are strong and yet lightweight structures that have already been successful for learning many tasks, such as bitrate control for real-time VVC [39]. Considering $x$ as the input vector to MLP, $\hat{y}$ as the predicted output, $h(x;\theta)$ as the hidden layer with parameters $\theta$, (8) describes the MLP architecture. Here, $\sigma(.)$ is the SoftMax function that bounds the outputs into the range [0-1], and $g$ represents the $tanh(.)$ as the activation function.

$$\hat{y} = \sigma\big(g(h(x;\theta))\big) \quad (8)$$

Fig. 6 (top-left) depicts the overall architecture of the offline model. It can be observed that the input to the network is the first modality feature vector, $x_1$, and the model learns to predict a vector of scores between 0 and 1 for each class, i.e., $\hat{y}_1$. The model uses a hidden layer with only 100 parameters, and is trained offline on a training set (details of training in 4.1). The number of parameters is found experimentally (similarly done for the online and the mixed models).

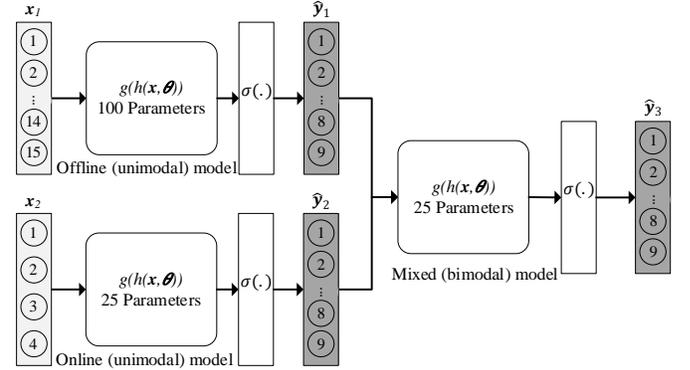

**Fig. 6.** Architecture of the three proposed MLP models

As observed in Fig. 6 (bottom-left), the online model receives the second feature modality, $x_2$, with only 4 features, and similar to the offline model, learned to predict a vector of scores between 0 and 1 for each class, i.e., $\hat{y}_2$. This model is trained online for each video scene, with a few samples from the first frame of each scene. Then the trained model is used to predict the best modes for the rest of this scene. The hidden layer has only 25 parameters which allows a fast online training.

The third model (Fig. 6–right side) is a mixed model that is trained on the predicted scores of the online and offline models. The model receives the concatenated scores of the first two networks as the input $x_3$, as in (9), and predicts a vector of scores, $\hat{y}_3$. Similar to the online model, this model is trained online on the first frame of each video scene, and is used to predict the best modes for the rest of the scene. A hidden layer with only 25 parameters is used to fuse the first two predictions and predict the final score.

$$x_3 = \hat{y}_1 || \hat{y}_2 \quad (9)$$

To train each of these three models, the cross-entropy loss is calculated as in (10), where $\mathbb{E}_{x \sim p} f(x)$ is the expectation of $f(x)$ with respect to $p(x)$. Cross-entropy measures the dissimilarity between the empirical distribution form the data ($p_{data}$), and the model distribution ($p_{model}$) [37]. For the purpose of this work, cross-entropy can be simplified into (11), where $y^{(i)}$ and $\hat{y}^{(i)}$ are the ground truth and the predicted scores for the $i^{th}$ class, respectively. This loss is minimized via the scaled conjugate gradient backpropagation algorithm (SCG) [40]. Details of the training process are discussed in section 4.1.

$$\mathcal{J}(\theta) = -\mathbb{E}_{x,y \sim p_{data}} \log p_{model}(y|x) \quad (10)$$

$$= -\sum_{i=1}^{9} y^{(i)} \log \hat{y}^{(i)} \quad (11)$$

$$\theta = argmin(\mathcal{J}) \quad (12)$$

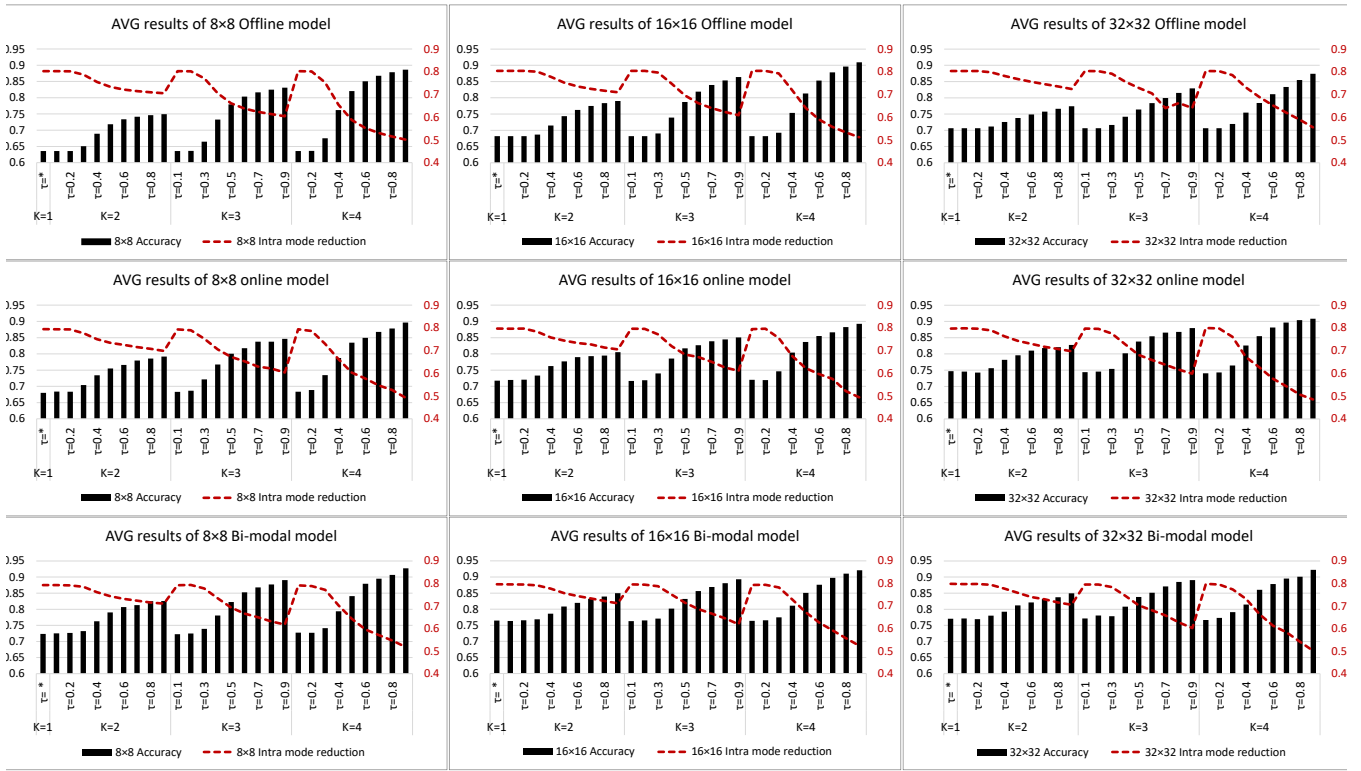
**Fig. 7.** Average accuracy and intra model reductions achieved using the three proposed methods, and different values of τ and K

### 3.4 Finding the best mode and measuring the accuracy

After training, each trained network returns an array of 9 scores for each input block, corresponding to the probability of each mode class. The easiest way to use these scores would be to choose the class with the highest score as the wining class, and checking the intra modes within that class only. However, for many blocks, a dominant edge direction does not exist, and the score of best mode is still rather low. This shows low confidence and suggests that more intra modes should be checked for this block to find the best mode.

Accordingly, a list of modes to be checked by the encoder is formed, to find the best mode, as in (13). If the mode class with the highest score, has a score higher than a threshold $\tau$, only the modes within this class will be checked. Otherwise, modes of $K$ classes with the highest scores are checked to find the best mode. Moreover, the DC and Planar modes are always checked.

$$\begin{cases} List = class\ i, & if\ i = argmax(\hat{y})\ and\ \hat{y}^{(i)} \geq \tau \\ List = K\ best\ mode\ classes, & else \end{cases} \quad (13)$$

To deeply analyze the model's accuracy, all three trained models were used to predict the best mode of 9 test video scenes, using different values of $\tau$ and $K$. Details of the training and test videos is discussed in section 4.1, and the online and mixed models are trained online per video scene.

Fig. 7 presents the accuracy and intra complexity reduction achieved for threshold values ($\tau$) from 0.1 to 0.9, and number of selected classes ($K$) from 1 to 4. Here, accuracy refers to the chance of the best mode being included in the list of tested modes. It can be observed that setting $K=1$ and $\tau=*$ (i.e., choosing only the class with the highest score) gives the lowest accuracy and the highest intra mode reduction. Increasing $K$ and $\tau$ increases the accuracy and reduces the achieved complexity reduction. It can be seen that increasing $\tau$ above 0.7 leads to very small improvement of accuracy, while reducing the complexity reduction. Also, the improvement for $K>3$ is not significant considering the reduction in the achieved complexity reduction. Hence, the two parameter settings of $K=2$, $\tau=0.7$ and $K=3$, $\tau=0.7$ are selected as the best trade-offs, and used for experiments. Moreover, it can be observed that using the mixed bimodal model considerably improves the accuracy compared to the online and offline models.

### 3.5 Fast VVC intra coding algorithm

This subsection describes the BLINC algorithm for fast VVC intra coding algorithm based on the three proposed learning models described in subsection 3.3. Algorithm 1 details this algorithm, receiving video frames of a scene, $F_0$ to $F_n$, and the learning algorithm described as the offline, online, or the mixed model. Furthermore, Fig. 8 sketches the coding timeline for each of the three methods used in this algorithm, for better understanding (not to scale actual timings. For timing refer to Table 6).

For the offline model, blocks of each frame ($F_0$ to $F_n$) are first processed using the offline trained model to find the list of candidate modes to check. The DC and Planar modes are also added to the list. Then, only the angular modes with an even index in the list (i.e., half of the modes) are checked via RMD. To cover the odd modes, the two modes around the best angular mode so far are checked. Finally, the $R$ modes with the lowest RMD cost are checked via RDO and the best mode is determined as the one with the lowest RDO cost.

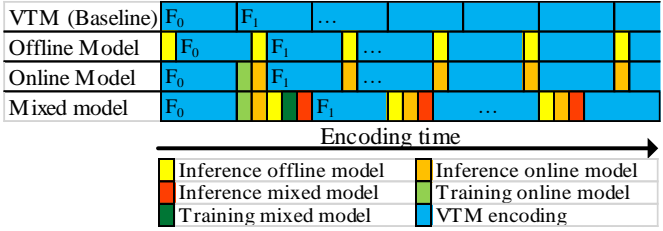

**Fig. 8.** Encoding timeline for the baseline VVC encoding, and the three methods described in Algorithm 1 (not to scale)

**Algorithm 1:** Fast VVC intra coding, BLINC, for each video scene
**Input**: video frames $F_0$ to $F_n$, learning model (offline, online, or mixed)
1  **if** (offline model)
2      jump to line 13
3  **else if** online model
4      encode $F_0$ with bassline VVC, collect $x_2$
5      train the online model on $x_2$ data
6      jump to line 13
7  **else if** mixed (bimodal) model
8      encode $F_0$ with bassline VVC, collect $x_2$
9      train the online model
10     find $x_3 = \hat{y}_1 || \hat{y}_2$ for $F_0$ using the offline and online models
11     train the mixed model on $x_3$ data
12 **end** if
13 **for** all blocks of video frames $F_0$ to $F_n$ do
14     **if** (offline or mixed model)
15         find $\hat{y}_1$ using the offline model
16     **end** if
17     **if** (online or mixed model)
18         find $\hat{y}_2$ using the online model
19     **end** if
20     **if** (mixed model)
21         find $\hat{y}_3$ using the mixed model
22     **end** if
23     find the list of candidate modes to check using (13)
24     add DC and Planar to the list
25     perform RMD on modes in list with even index
26     perform RMD on the best angular mode ±1
27     perform RDO on $R$ modes with the lowest RMD cost
28     encode with the mode with lowest RDO cost
29 **end** for

For the online model, the first frame, $F_0$ is first encoded with the baseline VVC encoding process to obtain the features of the second modality, $x_2$. Then the online model is trained on $x_2$. For the rest of the frames, $F_1$ to $F_n$, the coding process is similar to the case of offline model (previous paragraph), only using the online trained model instead of the offline one.

For the mixed (bimodal) model, after training the online model as described above, the scores $\hat{y}_1$ and $\hat{y}_2$ are obtained for $F_0$ using the offline and online trained models, respectively. Then the mixed model is trained on the concatenated scores, $x_3$. To encode blocks of $F_1$ to $F_n$, first the score on offline and online models for each block are derived. Then the scores are passed to the mixed model to find the final scores, which determines the list of models to check. The rest of the process is similar to the online and offline models described above.

## 4. Experimental Results

This section analyzes the performance of the proposed fast intra VVC coding method. The first subsection summarizes the experimental setup and test conditions. Then, the proposed learning strategies are analyzed and the accuracy of the methods are reported. Next, the encoding performance of the proposed method is assessed and compared with competing methods. Finally, a timing analysis is reported to measure the computational overhead of the proposed methods.

### 4.1 Experimental setup and configurations

Thirteen video sequences (each including one scene) from the VVC common test conditions [3] are used for training and test of the proposed method. Four sequences (Parkscene 1920×1080, BasketballDrive 1920×1080, FourPeople 1280×720, and RaceHorses 832×480) are used to extract training data samples for training of the offline model. Training samples are selected using four Quantization Parameters (QP) of 15, 25, 35, and 45 (different from testing QPs to be fair) to cover a wide range of video bitrates. Also, online and mixed models use the first frame of each test sequence for the online training. Nine different video sequences (refer to Table 3) are used only for testing.

For training, the SCG algorithm [40] was used with a maximum of 1000 epochs for the offline, and maximum 100 epochs for online and mixed models. The training stops if the validation loss does not improve for 6 consecutive validation checks. All training and inferences are done on a Central Processing Unit (CPU) and GPU is disabled to simulate a realistic processing platform where GPUs are not available.

The proposed models described in 3.3 are integrated into the VTM 7.0 [10]. To assess the performance of encoding, all encodings are repeated for the baseline VTM and the proposed methods, using the All Intra (AI) configuration [3], on the first 100 frames of the test videos, with QP values of 22, 27, 32, and 37. The Bjøntegaard Delta Rate (BD-Rate) [41] and the average Time Saving (TS) are calculated for each video scene.

All tests are performed on a modern machine using a single thread program compiled on a 64-bit Microsoft Visual Studio 2019 compiler, Intel Core i7 4650 processor with a working frequency of 4GHz, 8 GBs of main memory, and Microsoft Windows 10 platform.

### 4.2 Learning performance and accuracy

Fig. 9 (top) shows the learning loss (cross-entropy) per epoch for the offline model. As this model is trained offline once, it is trained for a maximum of 1000 epochs. It can be observed that the training loss almost saturates towards the end and the best validation loss, 0.1278, is achieved at epoch 925 and slightly rises afterwards.

For the online and mixed models, as the training samples and the network parameters are small, the training can be done with much fewer epochs (set to maximum 100 epochs). Fig. 9 bottom-left and bottom-right correspond to the case of BasketballDrill_1080 and Kristen&Sara_720 videos respectively, and compare the validation loss of online and mixed model, with the best loss of offline model. First, it is observed that both models reach their best loss after 27-41 epochs and stop as the loss does not improve anymore. Second, both losses overtake the best result of offline mode within very few epochs, and reach a much smaller loss over the time. For Kristen&Sara, the online and mixed models achieve 0.0879 and 0.0759 respectively, which are 31% and 40% smaller than the best loss of the offline model.

It is worth mentioning that the online and mixed models were also trained on 10 frames of each test video (instead of only the first frame) to see if the training performance improves due to larger training set. The test shows that such training for BasketballDrill and Kristen$arawill stop at epochs 172 and 187, and improves only by 4% and 3% compared to the case above, respectively. This shows that training on the first frame provides sufficient data and is a reasonable trade-off.

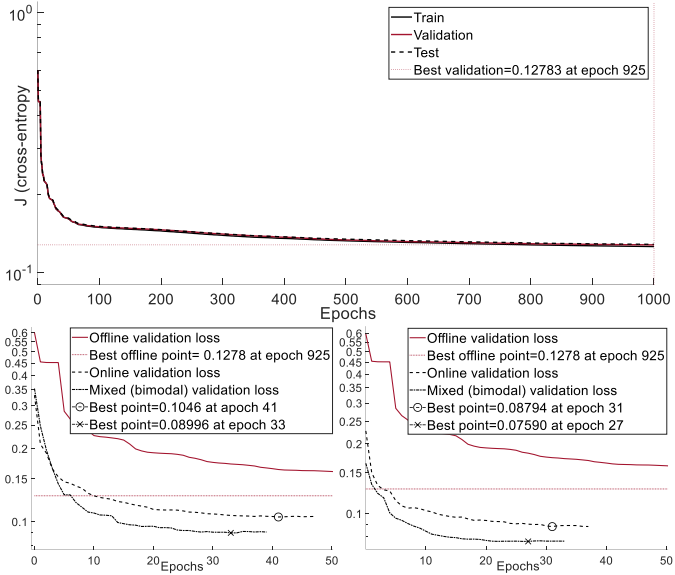

**Fig. 9.** Training loss (cross-entropy, logarithmic scale) of 16×16 models per epoch, for (top) offline, (bottom-left) online and mixed for BasketballDrill, and (bottom-right) online and mixed for Kristen&Sara

The accuracy of the trained models for different configurations are already presented and discussed in 3.4. Table 2 summarizes the accuracy of the two configurations selected for the encoding experiments, averaged over all of the test sequences. Similar to measurements in 3.4, accuracy refers to the chance of the best mode being included in the list of tested modes. It is observed that the offline model achieves a great performance by efficiently learning the blocks' texture. The online model can provide an even better accuracy compared to the offline, as it is trained per video scene. However, as discussed later in 4.3 its false predictions are more critical for compression, and achieves a worse coding performance. The mixed (bimodal) model however, mixes information from the texture (offline model) and the distribution of modes in the frame (online model) and achieves the best performance. Also, as this model does not directly use the neighboring blocks, this accuracy also translates into a great coding performance (see 4.3). Moreover, it is observed that the mixed model achieves a better improvement for smaller blocks, as they are more prone to noise. Finally, as the case with $K$=3 checks 3 best mode classes in case all classes are predicted with a $\tau$<0.7, it achieves a higher accuracy compared to $K$=2.

**Table 2.** Average accuracy (%) of each training model for the selected configurations

| Block | Configuration | Offline | Online | Mixed (bimodal) |
|---|---|---|---|---|
| 4×4 | $\tau$=0.7, $K$=2 | 73 | 78 | **82** |
|  | $\tau$=0.7, $K$=3 | 80 | 84 | **87** |
| 8×8 | $\tau$=0.7, $K$=2 | 74 | 78 | **81** |
|  | $\tau$=0.7, $K$=3 | 82 | 84 | **87** |
| 16×16 | $\tau$=0.7, $K$=2 | 77 | 79 | **83** |
|  | $\tau$=0.7, $K$=3 | 84 | 84 | **87** |
| 32×32 | $\tau$=0.7, $K$=2 | 76 | 82 | **83** |
|  | $\tau$=0.7, $K$=3 | 80 | 86 | **87** |
| 64×64 | $\tau$=0.7, $K$=2 | 84 | 87 | **88** |
|  | $\tau$=0.7, $K$=3 | 87 | 90 | **91** |

### 4.3 Encoding performance and complexity reduction

Table 3 reports the encoding performance of the proposed fast VVC intra coding with a number of RDO modes ($R$) of 2, based on the three training methods. Smaller BD-Rate and higher TS shows a better performance. It is observed that the offline method achieves on average 14.5% and 15.6% time reductions with only 1.23% and 1.61% BD-Rate increase, for $K$=3 and 2, respectively. Using $K$=3 often resulted in a better BD-Rate, while slightly reducing the achieved time saving.

With a BD-Rate increase of 1.62% and 2.08% and similar time savings as the other two methods for $K$=3 and 2, the online model achieves the worst results among the three models. This is explained by the fact that (1) the online model relies on the neighboring blocks and this can drift the mode decision towards the more probable modes, after some time, and (2) the model is oblivion to the texture characteristics and can fail near the boundary of two of more objects.

The mixed method achieves the best results among the three models, with an average of 1.17% and 1.47% BD-Rate increase and 14.7% and 16.1% time saving, for $K$=3 and 2, respectively. As this model benefits from both data modalities, it achieves a performance above both offline and online models, and achieves good performance in corner cases where the other two fail, such as Cactus and BasketballDrill.

To have a more thorough evaluation, Table 4 summarizes the average coding performance of the proposed method for different number of RDO modes ($R$), different training methods, and $K$=2 and 3. It is observed that with the increase of $R$ and $K$, BD-Rate increases and the time reduction decreases, as expected. The best compression efficiency is achieved for the mixed model with $R$=3, $K$=3, with a BD-Rate increase of 1.04% and complexity reduction of 6.9%. The mixed model with $R$=1, $K$=3 achieves a BD-Rate increase of 1.47% with a time saving of 23.5%, which is among the best time savings. Also, the reported BD-PSNRs are very low and show that the proposed method achieves almost similar quality as VTM, given the same bitrates.

To compare the proposed method with competing techniques, the method in [17] was reimplemented for VVC and tested with similar configurations as this work. Table 5 compares this work with two comparable configurations of this work. It can be observed that both the mixed model and offline model (similar to [17]) with $K$=3, $R$=1 significantly overtake [17] in both BD-Rate and time saving.

**Table 3.** Coding performance (BD-Rate (%)) and encoding Time Saving (TS (%)) for each method, for R=2 and τ=0.7

| | Offline | | | | Online | | | | Mixed (bimodal) | | | |
|---|---|---|---|---|---|---|---|---|---|---|---|---|
| | K = 3 | | K = 2 | | K = 3 | | K = 2 | | K = 3 | | K = 2 | |
| | BD-Rate | TS | BD-Rate | TS | BD-Rate | TS | BD-Rate | TS | BD-Rate | TS | BD-Rate | TS |
| Cactus 1080 | 2.62 | 18 | 2.95 | 19 | 1.54 | 15.9 | 1.87 | 16.7 | 1.45 | 15.3 | 1.82 | 17.9 |
| BQTerrace 1080 | 0.9 | 14.2 | 1.19 | 15.2 | 1.13 | 14 | 1.44 | 15.1 | 0.69 | 13.6 | 0.91 | 14.9 |
| BasketballDrill 1080 | 2.01 | 11.1 | 3.04 | 12.9 | 2.2 | 12.5 | 3.05 | 14.3 | 1.57 | 11.4 | 2.13 | 12.5 |
| Kimono1 1080 | 0.33 | 16.5 | 0.46 | 17.9 | 0.64 | 19.6 | 0.77 | 20.4 | 0.49 | 18.8 | 0.55 | 19.7 |
| Johnny 720 | 1.29 | 13.8 | 1.73 | 14.6 | 2.1 | 14.1 | 2.72 | 15.3 | 1.18 | 13.4 | 1.54 | 15.3 |
| Kristen&Sara 720 | 1.23 | 13.6 | 1.75 | 14.4 | 2.3 | 13.5 | 2.95 | 15.4 | 1.21 | 13.9 | 1.6 | 15.1 |
| MobileCalendar 720 | 0.59 | 15 | 0.78 | 16.4 | 1.62 | 16.9 | 2.12 | 18.6 | 1.98 | 18.2 | 2.3 | 18.7 |
| PartyScene 480 | 0.93 | 14.2 | 1.14 | 14.9 | 1.28 | 15 | 1.56 | 16.4 | 0.84 | 14.2 | 1.04 | 15.5 |
| BQMall 480 | 1.17 | 14 | 1.47 | 14.9 | 1.74 | 13.8 | 2.18 | 15.7 | 1.09 | 13.4 | 1.36 | 14.9 |
| Average | 1.23 | 14.5 | 1.61 | 15.6 | 1.62 | 15 | 2.08 | 16.4 | 1.17 | 14.7 | 1.47 | 16.1 |

**Table 4.** Coding performance (BD-Rate (%) and BD-PSNR (dB)) and encoding TS (%) for different number of RDO (R), for τ=0.7

| | Offline | | | | | | Online | | | | | | Mixed (bimodal) | | | | | |
|---|---|---|---|---|---|---|---|---|---|---|---|---|---|---|---|---|---|---|
| | K = 3 | | | K = 2 | | | K = 3 | | | K = 2 | | | K = 3 | | | K = 2 | | |
| RDO | BD-Rate | BD-PSNR | TS | BD-Rate | BD-PSNR | TS | BD-Rate | BD-PSNR | TS | BD-Rate | BD-PSNR | TS | BD-Rate | BD-PSNR | TS | BD-Rate | BD-PSNR | TS |
| R=1 | 1.52 | -0.07 | 23 | 1.89 | -0.08 | 24.1 | 1.9 | -0.09 | 23.4 | 2.34 | -0.11 | 24.9 | 1.47 | -0.07 | 23.5 | 1.75 | -0.08 | 24.4 |
| R=2 | 1.23 | -0.05 | 14.5 | 1.61 | -0.07 | 15.6 | 1.62 | -0.08 | 15 | 2.08 | -0.1 | 16.4 | 1.17 | -0.06 | 14.7 | 1.47 | -0.07 | 16.1 |
| R=3 | 1.11 | -0.05 | 7.3 | 1.48 | -0.07 | 8.1 | 1.51 | -0.07 | 7.1 | 1.93 | -0.09 | 8.9 | 1.04 | -0.05 | 6.9 | 1.33 | -0.06 | 8.4 |

**Table 5.** Coding performance comparison with competing method

| | [17] Reimplemented for VVC | | | Offline, R=1, K = 3 | | | Mixed, R=1, K = 3 | | |
|---|---|---|---|---|---|---|---|---|---|
| | BD-Rate | BD-PSNR | TS | BD-Rate | BD-PSNR | TS | BD-Rate | BD-PSNR | TS |
| Cactus 1080 | 2.53 | -0.08 | 20.5 | 2.89 | -0.1 | **27** | **1.67** | -0.06 | 25.1 |
| BQTerrace 1080 | 2.16 | -0.1 | 18.8 | 1.17 | -0.06 | **22.9** | **0.97** | -0.05 | 22.7 |
| BasketballDrill 1080 | 4.25 | -0.19 | 17.3 | 2.64 | -0.12 | **19.3** | **2.11** | -0.09 | 19.2 |
| Kimono1 1080 | 0.57 | -0.02 | 21.4 | **0.37** | -0.01 | 23.4 | 0.57 | -0.02 | **26.6** |
| Johnny 720 | 3.45 | -0.13 | 17.5 | **1.49** | -0.05 | **21.3** | 1.58 | -0.06 | 21.2 |
| Kristen&Sara 720 | 3.15 | -0.15 | 18.1 | 1.61 | -0.07 | **21.6** | **1.59** | -0.07 | 21.5 |
| MobileCalendar 720 | 1.2 | -0.07 | 20.5 | **0.74** | -0.04 | 25 | 2.14 | -0.13 | **28.4** |
| PartyScene 480 | 1.43 | -0.1 | 19.6 | 1.23 | -0.09 | 23.8 | **1.15** | -0.08 | **24.3** |
| BQMall 480 | 2.2 | -0.12 | 19 | 1.5 | -0.08 | 22.4 | **1.46** | -0.08 | **22.8** |
| Average | 2.33 | -0.11 | 19.2 | 1.52 | -0.07 | 23 | **1.47** | -0.07 | **23.5** |

Finally, Fig. 10 compares the percentage of intra mode reductions achieved using different learning methods and different configurations. It is observed that the mixed model achieves a better performance compared to both online and offline methods with similar configuration. The main reason is that, using both data modalities, the mixed model more often reaches a score with high confidence for the best mode (i.e., a higher τ), and can more often avoid checking two or more mode classes.

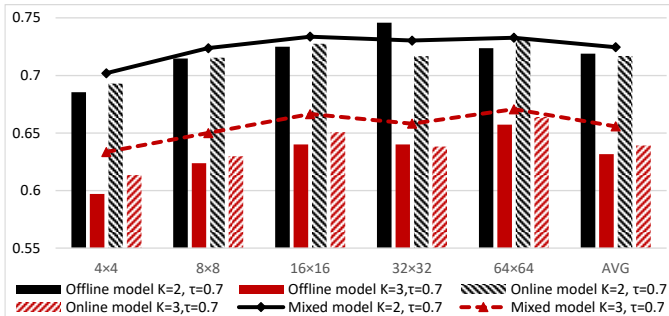

**Fig. 10.** Comparison of intra mode reductions (%) achieved using different configurations and learning methods

### 4.4 Timing analysis and implementation issues

One of the advantages of the proposed method, compared to the competing methods, is its low computational overhead, which makes it more practical. To assess this, Table 6 reports the time complexity of each coding component, compared to the coding time of a single frame, and the coding time of the whole video (100 frames) encoded using AI configuration, as in experiments of 4.3. It should be noted that VTM's AI configuration, encodes only one in every eight frames to save time (given that intra frames are coded independently). Thus, encoding each scene includes the encoding times of 13 video frames. These components correspond to the components of the timeline in Fig. 8.

Firstly, it can be observed that the inference times for all three models is negligible compared to the encoding time. Second, for online and mixed models, the online training times are only 0.06% and 0.04% of the encoding time, respectively, confirming their lightweight training property. Thirdly, for the mixed model (i.e., the most complex one) the total overhead time to encode the whole video, is only 0.2%, making it practical for real-world applications.

**Table 6.** Time overhead of each component in the proposed mixed method, compared to the coding time of whole video (100 frame AI), and a single frame

| | Time (%) in total video | | Time (%) in one frame | |
|---|---|---|---|---|
| | BasketDrill | Kris&Sara | BasketDrill | Kris&Sara |
| Inference offline (1 frame) | 0.003 | 0.003 | 0.039 | 0.042 |
| Inference online (1 frame) | 0.002 | 0.002 | 0.028 | 0.024 |
| Inference mixed (1 frame) | 0.002 | 0.003 | 0.029 | 0.033 |
| Training online | 0.06 | 0.08 | 0.781 | 1.044 |
| Training mixed | 0.039 | 0.061 | 0.513 | 0.795 |
| Total Overhead for mixed | 0.2 | 0.245 | - | - |
| Total VVC Enc | 99.8 | 99.755 | - | - |

To have a better view of the timing overhead, Table 7 compares the overhead percentage, number of features used for intra mode prediction, and the number of learnable parameters, among the proposed methods (bimodal and unimodal) and two alternative approaches. The methods presented in [11] (also used in [31]) use a signal processing solution to handcraft features from the DT-CWT which are then learned using SVM. The method in [12] on the other hand, uses a CNN to predict the best mode. Both methods are re-implemented, trained offline, and tested in similar conditions as this work. It is observed that the bimodal and unimodal (offline) methods have a negligible overhead of 0.2% and 0.04%, respectively. The DT-CWT based method has a larger overhead of 1.47%. The CNN method has the largest overhead, which is 33.5%. This complexity is expected as this CNN has 650K parameters which results in a huge computational cost for each inference. While CNNs can speed up on a GPU, it should be noted that (1) GPUs are not available for all video applications/platforms, (2) while a GPU implementation improves the timing, the power and energy overhead would still be huge, (3) even with a GPU, the overhead will still be high due to the large number of parameters.

**Table 7.** Comparison of overhead and learnable parameters among different approaches

| Method | Overhead time (%) | Number of features | Number of Parameters |
|---|---|---|---|
| Mixed (bimodal-BLINC) | 0.2 | 15+4 | 150 |
| Offline (unimodal-BLINC) | 0.04 | 15/4 | 100/25 |
| Signal Proc-based [11] | 1.47 | 5 | 5 |
| CNN-based [12] | 33.5 | 65K | 650K |

Finally, the proposed method can be adopted with a low implementation cost in a hardware or software platform. The offline method would only require the MLP and integration of the learned parameters to run, which takes very low cost. The online and mixed models also need the ability to train the MLP online. This can easily be done in a software platform, but is harder in hardware-based implementations. The method in [11] uses SVM which is medium cost, but for feature extraction the DT-CWT is required which adds a significant implementation cost. The CNN-based approach [12] requires the CNN architecture which is easy to integrate in software, but very costly for a hardware implementation. Moreover, even for a software implementation, the CNN cannot run in real-time without GPU assistance.

## 5. Conclusion

This paper proposed a lightweight machine learning algorithm to simplify the VVC intra coding. The proposed method, BLINC, uses the DCT coefficients and neighboring modes to extract two data modalities for the intra mode decision task. Then, a two-step feature reduction method was proposed to efficiently reduce the feature size, which allows learning the task with a lightweight model. Three training strategies namely, offline, online, and a mixed models were proposed that learn the intra mode decision task using the first, second, and both data modalities. Experimental results confirm the performance and effectiveness of the proposed method.

Some important conclusions were made based on the experiments in this work:

- The first proposed feature modality uses the DCT coefficients of intra modes, which is already part of any VVC encoder; hence, it imposes almost zero overhead. It was illustrated how alternative solutions impose multiple times more computational overhead.
- It was shown that using the intra mode distribution in the frame (neighboring modes) as the feature (the second modality), can lead to a good modeling accuracy. However, the corner cases where the model fails, are more critical in terms of coding efficiency and leads to higher compression loss, compared to the offline and mixed models.
- It was demonstrated that using both data modalities (bimodal model) in a mixed online-offline way, can improve the performance above both unimodal methods as it covers both models' corner cases.
- It was observed that using the proposed feature reduction scheme, which is specially designed for this task, can enable training on lightweight models. Such models can run on any processing unit, which is in contrast with deep CNN based methods that perform well only on GPUs.

The proposed models can be deployed based on the requirements. The offline unimodal model provides very good results and can be deployed on any platform with minimum costs and overheads. The mixed bimodal model, provides the best results and can be deployed easily on any software/hardware platform, only adding the cost of implementing online training on top of an existing encoder.